\documentclass[pre,aps,twocolumn,showpacs]{revtex4-1}

\draft 
\usepackage{graphicx}

\begin{document}


\title{Randomness versus specifics for word-frequency distributions} 



\author{Xiaoyong Yan$^{1,2}$ and Petter Minnhagen$^{3}$}
\email[]{Petter.Minnhagen@physics.umu.se}
\affiliation{
$^1$Systems Science Institute, Beijing Jiaotong University, Beijing 100044, China\\
$^2$Big Data Research Center, University of Electronic Science and Technology of China, Chengdu 611731, China\\
$^3$IceLab, Department of Physics, Ume{\aa} University, 901 87 Ume{\aa}, Sweden
}

\date{\today}

\begin{abstract}

The text-length-dependence of real word-frequency distributions can be connected to the general properties of a random book. It is pointed out that this finding has strong implications, when deciding between two conceptually different views on word-frequency distributions, \textit{i.e.} the specific `Zipf's-view' and the non-specific `Randomness-view', as is discussed. It is also noticed that the text-length transformation of a random book does have an exact scaling property precisely for the power-law index $\gamma=1$, as opposed to the Zipf's exponent $\gamma=2$ and the implication of this exact scaling property is discussed. However a real text has $\gamma>1$ and as a consequence $\gamma$ increases when shortening a real text. The connections to the predictions from the RGF(Random Group Formation) and to the infinite length-limit of a meta-book are also discussed. The difference between `curve-fitting' and `predicting' word-frequency distributions is stressed. It is pointed out that the question of randomness versus specifics for the distribution of outcomes in case of sufficiently complex systems has a much wider relevance than just the word-frequency example analyzed in the present work. 
 
\end{abstract}


\maketitle 

\section{Introduction} \label{sec:1}

The question of trying to understand what \textit{linguistic} information is hidden in the \textit{shape} of the word-frequency distribution has a long tradition. It goes back  to the first part of the twentieth century when it was discovered that the word-frequency distribution of a text typically has a broad ``fat-tailed" shape, which often can be well approximated with a power law over a large range \cite{estroup16,zipf32,zipf35,zipf49}. This led to the empirical concept of Zipf's law which states that the number of words that occur $k$-times in a text, $N(k)$, is proportional to $1/k^2$ \cite{zipf32,zipf35,zipf49}. The question is then what special principle or property of a language causes this power law distribution of word-frequencies and this is still an ongoing research \cite{mand53,li92,baayen01,cancho03,mont01,font-clos2013}.

In middle of the twentieth century Simon in Ref. \cite{simon55} instead suggested that, since quite a few completely different systems also seemed to follow Zipf's law in their corresponding frequency distributions, the explanation of the law must be more general and stochastic in nature and hence independent of any specific information of the language itself. Instead he proposed a random stochastic growth model for a book written one word at a time from beginning to end. This became a very influential model and has served as a starting point for much later works \cite{kanter95,doro01,zanette05,masucii06,cattuto06,lu13}. In the `Simon-view' the shape of the word-frequency distribution does not reflect any specific property of a language but is shaped by a random stochastic element. An extreme random model was proposed in the middle of the twentieth century by Miller in Ref. \cite{miller57}: the resulting text can be described as being produced by a monkey randomly typing away on a typewriter. However the properties of the monkey book are quite unrealistic and different from a real text \cite{bern11b}. This `Randomness-view' was recently developed further in a series of paper in terms of concepts like Random Group Formation, Random Book Transformation and the Meta-book \cite{bern11b,bern10,bern09,baek11,yan14}. A crucial difference, compared to the `Zip´s-view', is that the `Randomness-view' is based on the notion that the shape of the word-frequency distribution is a general consequence of randomness which carries no specific information of the language. 

In other words the ´Zipf-view´ is leaning more on the idea that a language is a special system and that as a consequence the functional form of the word-frequency distribution reflects some specific property of the language, whereas the ´Randomness-view´ maintains that very little specific language information can be extracted from this distribution.

The concept of randomness in a text dates back to at least 1913 and A. Markov \cite{markov1913,hayes2013}: Markov demonstrated that even an exquisitely crafted poem like Pushkin's Eugene Onegin, when viewed as a string of letters, contained random features like e.g. how often a randomly chosen letter is followed by a consonant  or a vowel. This was at the beginning of what developed into the fundamental statistical concept of Markov chains. This begs the conceptual question of how something crafted with such an amount of intention, purpose and meaning could possibly contain something entirely random. A somewhat related question is hidden within the decimal tail of the number $\pi=3.14159265.....$: The decimal tail of  $\pi$ has a definite cause since it is the ratio between the circumference and diameter of a circle. Thus every decimal in the expansion is solidly given. However, if you pick a decimal place randomly and read  off its value and ask yourself what the value of the next decimal might be, then it is with equal probability any of the numbers 0,1,..,9. Thus the poem  Eugene Onegin and the number $\pi$ both display some randomness in spite of their perfectly deterministic cause.

From a statistical point of view the decimal tail of $\pi$ is pseudo-random and equivalent to a number-series created by throwing a dice with ten fair outcomes. However, if the only thing you know is that the decimal tail of $\pi$ is equivalent to a pseudo-random series, throwing the dice will not give you any information as to the ratio between the circumference and the diameter of a circle.

Words in a text are random in an  analogous  fashion; A specific word occurs $k$ times in the text and $N(k)$ specific words occur the same number of times. Suppose you randomly pick a word in the text and that this word occurs $k^{'}$ times. What is the total number of occurrence in the text of the following word? The randomness view argues that this occurrence is random and given by a probability proportional to $N(k)$.  The dice $N(k)$ itself can be estimated using the maximum entropy principle \cite{baek11}.

The fact that  frequency distributions of possible outcomes for some sufficiently complex deterministic systems reduce to equivalent random distributions is not restricted to words\cite{baek11,lee12,baek11b,bokma13,baz10,yan13}. Deterministic systems which display random features are termed \textit{pseudo-random}. In the discussion section some more examples are mentioned. However, in the present paper we analyze the consequences for  words in a text. The general point is that the ideas of scale-freeness ingrained into the various Zipf's law approaches are superseded by the inherent randomness, which we argue is a very basic property of a written text.

In order to be concrete we will focus on the difference between, on the one hand, a generalized scaling law for word-frequency distributions proposed by Font-Clos et al in Ref. \cite{font-clos2013} and suggesting a bona fide specific property of a language, and, on the other hand, the general predictions from the ´Randomness-view´ \cite{bern11b,bern10,bern09,baek11,yan14}. 

We will in the present paper use the following notation: $N_M(k)$ ($N_M(\geq k)$) is the number of distinct words which occur $k$-times ($k$-times or more) in a text which in total contains $M$ words. The scaling law proposed in Ref. \cite{font-clos2013} can be cast into the form 
$N_M(\geq k)=G(k/M)$.

In section \ref{sec:2}, we first demonstrate directly from raw data that $N_M(\geq k)$ does indeed change shape with text-length in a very systematic manner such that the proposed scaling-form $N_M(\geq k)=G(k/M)$ cannot be conceptually valid. This means that this scaling function cannot be a true specific feature of the word-frequency distribution. In section \ref{sec:3}, we then compare the systematic length dependence of $N_M(\geq k)$ with the predictions from the `Randomness-view' and indeed find consistent agreement. We elucidate just how little information you need about the language in order to \textit{predict} the characteristic features of the data for the word-frequency. This has a crucial and more far reaching consequence: whenever you need very little information to describe a particular feature, then indeed very little specific information about the system can be extracted from this characteristic feature. In section \ref{sec:4}, we discuss and show that for a distribution $N_M(k) \propto 1/k$ the shape is indeed length-invariant under the randomness (more precisely under the Random Book Transformation assumption \cite{bern10,bern09,baek11,yan14}).  In Ref. \cite{bern09} it was observed that the limit of a very large text by an author seems to approach the limit
$N_M(k) \propto 1/k$. This suggests that this approximate scaling should work better the longer the text is. Some concluding remarks are added in section \ref{sec:5}, in particular on the applicability of the `Randomness view' to a much broader spectrum of complex systems.

\section{Scaling or no Scaling?} \label{sec:2}
The first issue is the factual situation. Does or doesn't the word-frequency distribution, $N_M(k)$ change shape when shortening the text-lengths $M$? Note that, in the present context, two curves have the same shape provided their log-log-plots can be slid on top of each other, so that one is entirely on top of the other.

\begin{figure*}
\includegraphics[width=0.8\textwidth]{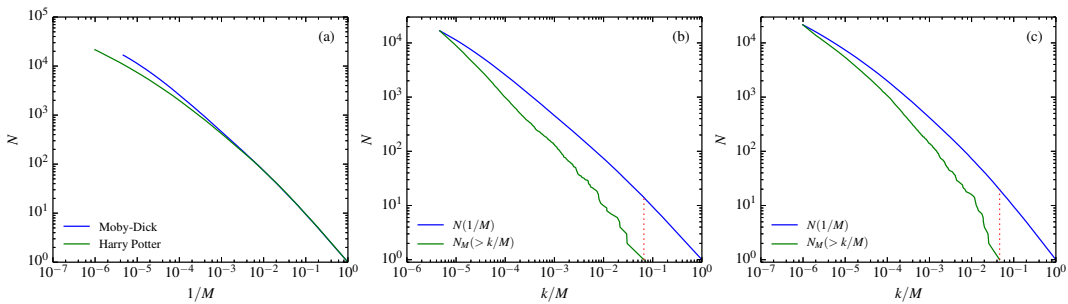}
\caption{Direct test of the scaling-relation $N_M(\geq k/M)=G(k/M)$: (a) shows $N(1/M)$, the number of distinct words, $N$, as a function of the inverse text-length, $1/M$ for Moby Dick (upper curve) and Potter 1-7 (lower curve). These two curves are on top for $1/M=1$ because any text of unit length contains precisely one word. However, for longer text-lengths they start to deviate. As explained in the text, the shape of $N(1/M)$ is a characteristics of the author and typically differs between authors (in this case Melville and Rowling: Rowling uses fewer specific words).
(b) $G(x)=N_M(\geq k/M)$ for fixed $M$ and varying $k$ is, in case of Moby Dick, compared to $G(x)=N_M(\geq k/M)$ for varying $M$ and fixed $k=1$. According to the scaling relation these two curves should be identical. The lower curve is $G(x)$ for fixed $M$ varying $k$ and the upper curve $G(x)$ for $k=1$ varying $M$. Note that in this latter case $G(x)=N(1/M)$, which is the upper curve in Fig. \ref{fig:1}(a). The dashes vertical line is the log of the text length which on the average contains one $the$ (=the most frequent word in the full text). This text length is about 15 words. This means that the two curves in Fig. \ref{fig:1}(b) by definition agree at the left end point, but have to differ by the log(15) at the right end point of the lower curve. So they are conceptually two distinct curves, which can never be connected by a scaling relation; (c) illustrates the same features as Fig. \ref{fig:1} (b) but for Potter 1-7. In this case the average text-length which contains on the average one $the$ is 23 words.
In Fig. \ref{fig:A1} $N_M(k)$ and $N(M)$ are replotted for Potter 1-7 in order to compare with power-law concepts like Zipf's law and Heaps' law.
}      

\label{fig:1}
\end{figure*}

Figure  \ref{fig:1}(a) gives a first illustration: the number of distinct words in a text, $N$, increases with the total number of words in the text $M$. $N$ as a function of $1/M$ is determined for two novels, \textit{Moby Dick} by H. Melville and \textit{Harry Potter 1-7} by J.K. Rowling, by taking averages over fixed text-length $M$. The data is plotted as $N$ against $1/M$ in a log-log scale. In the limit $M=1$ is $N=1$ (the first word is always a distinct word), so that the curves in Fig. \ref{fig:1}(a) overlap at this point (which is the right-most point in Fig. \ref{fig:1}(a)). However as $M$ increases the two curves start to systematically deviate. This demonstrates that these two curves do have different shapes. What conclusions can be drawn from this fact? The first conclusion is that the function $N(1/M)$ is \textit{not} a universal function of a language. Other possibilities are a) a property of a given language at a particular time-period (a language evolves slowly with time\cite{wijaya11,petersen12}), b) a unique property of an author writing a text or c) just a property of a particular text. As argued in Ref. \cite{bern09}, where the complete production of three different authors are compared, it is to good approximation a property of the author. In Ref. \cite{bern09} it was argued that this is related to the concept of a Meta book for an author, reminiscent of an author-fingerprint.

\begin{figure}
\includegraphics[width=0.5\textwidth]{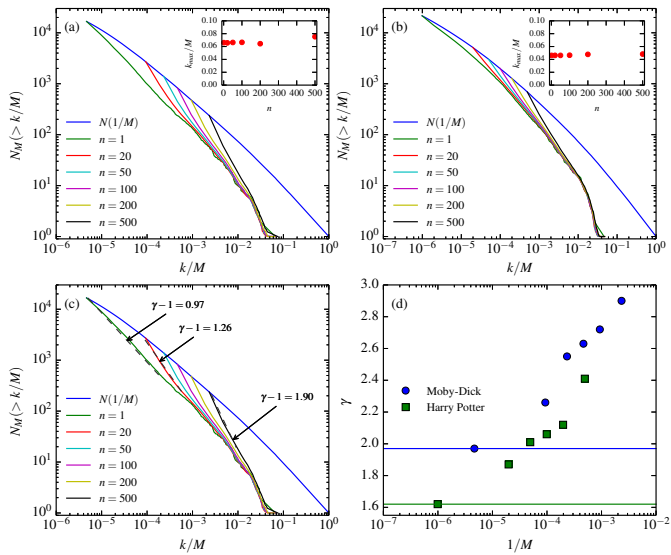}
\caption{Test of the scaling form $G(x)=N_M(\geq x=k/M)$ by comparing different text-lengths: (a) data for parts of Moby Dick, where $n$ denotes the average over an n$^{th}$-part of the novel. All these n$^{th}$-part curves to good approximation starts from the same right-most point because $k_{max}/M$ is to good approximation independent of the text length, as shown by the inset. However all curves by definition ends on the corresponding $N(1/M)$-curve (same curve as in Fig. \ref{fig:1}(b)). Clearly all these n$^{th}$-part curves have different shapes and are not connected by a scaling relation; (b) same as Fig. \ref{fig:2}(a) for Potter 1-7. 
(c) power law approximation for the n$^{th}$-parts for Moby Dick: $\gamma$ systematically increases with shortening the text-length.; (d) shows this systematic increase of $\gamma$ for Moby Dick and Potter. Note that $\gamma$ diverges, as the limit text-length corresponding to one $the$ on the average is approached (compare Figs. \ref{fig:1} (a) and (b)). The length-independent $\gamma$  predicted by invoking a scaling relation corresponds to the horizontal lines in Fig. \ref{fig:2}(d).}     

\label{fig:2}
\end{figure}

Next we investigate the number of distinct words, $N_M(\geq k)$, which occur more or equal to $k$ times in a text of length $M$. First one may note that since any distinct word must occur at least one time it follows that $N_M(\geq 1)=N(M)$. This means that if we instead change the number of occurrences $k$ to the relative number of occurrences $k/M$, then $N_M(\geq k)$ becomes $N_M(\geq k/M)$ and $N_M(\geq 1/M)=N(1/M)$. This latter form should, according the scaling form $N_M(\geq k)=G(k/M)$ be scale-invariant. To check this we again start with Moby Dick. First one notes that $N_M(\geq k_{max}/M)=1$, because the most common word `the' is a single word. This means that if one plots $N_M(\geq k/M)$, as a function of $k$, in the same plot as $N$, as a function of $1/M´$, then these two curves will coalesce at the left end points by definition. Next one takes the occurrence-randomness of the words in a text into account, which means that the ratio between the number of `the' and the total number of words $M'$ is to good approximation constant ($\approx 0.066$ for Moby Dick (see inset in Fig. \ref{fig:2}(a) and also Fig. 3 in Ref. \cite{bern09}). This means that, on the average, approximately every $15^{th}$ word in the text is a `the'. Or, in other words, a text-part of length $M_1=M/k_{max}$ on the average contains one `the', which means that $N(1/M_1)\approx 15$ for Moby Dick. Since  $N_M(\geq (k=k_{max})/M)=1$ is not equal to $N_M(\geq (k=1)/(M/k_{max})\approx15$, this shows that $N_M(\geq k/M)$ is not a scaling function in the complete range of the variable $k/M$. This inequality is illustrated in Fig. \ref{fig:1}(b) for Moby Dick and in Fig. \ref{fig:1}(c) for Potter 1-7. However, as explained above, the randomness view implies that it is a valid scaling for the most frequent word so that  $N_M(\geq (k=k_{max})/M)=1$ irrespective of $M$. The randomness view implies that this is the only point where the scaling is expected to be strictly valid \cite{bern11b,bern09}.

 In Fig. \ref{fig:2} we investigate this discrepancy in more detail: the novels are divided into text-lengths of given sizes $M$ and the $N_M(\geq k/M)$ is obtained as the average over such fixed text-lengths. A first observation is that $N_M(\geq 1/M)=N(1/M)$, which means that the left-most point for each text-length falls on the respective $N(1/M)$ curve shown in Fig. \ref{fig:1}. Fig. \ref{fig:2}(a) shows the result for Moby Dick. The complete Moby Dick contains $M=212 473$ words and corresponds to the lowest curve in Fig. \ref{fig:2}(a). The texts parts correspond to $M/10$, $M/100$, $M/500$, $M/1000$, and $M/3000$, respectively. One notes that all these text parts to good approximation starts from the same right-most point $k_{max}/M$. The reason for this is the following: the most frequent word in an English text is the word \textit{the}. To good approximation the density of \textit{the}'s is independent of where in the novel you are. In Moby Dick the density of \textit{the} is $k_{max}/M=0.066$ and is to good approximation constant and independent of the text lengths (compare inset in Fig. \ref{fig:2}(a)) \cite{bern09} .

\textit{If} the scaling $N_M(\geq k)=G(k/M)$ was entirely correct, \textit{then} all the data in Fig. \ref{fig:2}(a) should fall on the full $N(1/M)$ (highest curve in Fig. \ref{fig:1}(a)). This is clearly not the case. Next you can ask if the data for the parts of Moby Dick can be slid so that the overlap with the data for the full Moby Dick. However this is not possible because the curves describing the data do in fact have different shapes. Thus the scaling $N_M(\geq k)=G(k/M)$ is very approximate and limited to values close to $k_{max}$ (see Fig. \ref{fig:A3}(a) in Appendix \ref{apdx} for a estimate of relative errors). 
Figure \ref{fig:2}(b) contains the same analysis as Fig. \ref{fig:2}(a) but for the Harry Potter novels 1-7. Combined into one text these novels contain $M=1012790$ words and is hence about five times larger than Moby Dick. However, the conclusions are just the same as for Moby Dick. Yet, one notes that the full Harry Potter and a twentieth-part of Harry Potter, for larger values of $k/M$, overlap to good approximation in Fig. \ref{fig:2}(b). We will come back to the issue of what might be implied by this particular overlap.

As pointed out in Ref. \cite{font-clos2013}, the implication from the scaling function is that, if the full text can be approximated by a power law, then all the text-parts should be well approximated with the same power-law. In other words the implication is that the power-law index $\gamma$ does not change with text-size, in direct contradiction to the prediction from the `Randomness-view' \cite{bern09,bern11b}. Figure \ref{fig:2}(c) shows that the full Moby Dick can indeed be approximated by a straight-line (over a range, see Appendix \ref{apdx}). The slope of this line is $\gamma-1=0.97$. However, also the text-parts can to good approximation be approximated by such lines, but these lines become steeper the smaller the text part. Thus the power law index $\gamma$ does systematically increase with smaller text size. Figure \ref{fig:2}(d) shows this systematic increase in the power-law index with decreasing text length. Note that as the text-length limit $M_1=M/k_{max}$ is approached $\gamma$ diverges (see Figs. \ref{fig:1}(b) and (c)).  The prediction, based on invoking the scaling assumption\cite{font-clos2013}, is that $\gamma$ is constant (the horizontal line in Fig. \ref{fig:2}(d)) which has no support by the data.

In this section we have demonstrated that the scaling law $N_M(\geq k)=G(k/M)$ is not borne out by data. This conclusion was reach by carefully analyzing the consequences of such a scaling law, rather than just trying to fit it to the data over a limited range. Before discussing the possible implications, we will in the next section widen the perspective and describe what the `Randomness-view' predicts and implies.

\section{Randomness-predictions} \label{sec:3}

\begin{figure}
\includegraphics[width=0.5\textwidth]{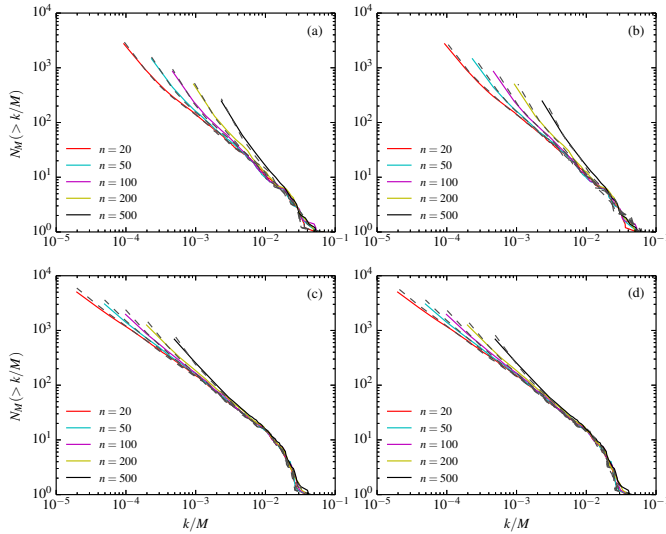}
\caption{Randomness predictions for partitions of texts. (a) shows the $n^{th}$-parts of Moby Dick (full curves) and the randomness prediction following from the homogeneous assumption (dashed curves). The near perfect agreement suggests that world-frequency distributions for the $n^{th}$-parts follow from simple statistics. (b) goes one step further. Here the starting knowledge is just $M$, $N$, and $k_{max}$ for the full text of Moby Dick. The $n^{th}$-parts are predicted by first using randomness in the form of the maximum entropy principle and RGF to obtain $P(k)$ for the full text and then subsequently using randomness in the form of RBT-transformation (Eqs (\ref{eq_RBT1},\ref{eq_RBT2},\ref{eq_RBT3}) to get the parts. The $n^{th}$-parts predictions are again given by the dashed curve and the agreement is nearly as good as in (a) (see Appendix \ref{apdx} for a plot of relative errors). This suggests that the shapes of word-frequency distributions contain basically no explicit linguistic information. (c) and (d) give the corresponding results for the Harry Potter data.}
\label{fig:3}
\end{figure}

The `Randomness view' of word-frequency distributions is based on two assumptions. The first is that a text written by an author is homogeneous (i.e. the chance of randomly picking a word of occurance $k$ is equal over the text) and the second randomness assumption enters through the maximum entropy principle. More precisely the first means that if you enumerate the word positions, $M$, in the text by $i=1,..M$ then the probability to find a word which occurs $k$ times in the text is to good approximation independent of the position $i$ within the text. For example you do not find more rare words at the end of the text than in the beginning and the most common word `the' is to good approximation evenly spread through the text. This assumption can be expressed by a mathematical transformation, the Random Book Transformation(RBT) \cite{bern09,bern10}, which will be described and discussed below in the present section. However, the basic consequence of the homogeneous assumption is that if you take an n$^{th}$-part of the text the \textit{word-frequency distribution} is to good approximation the same as if you just randomly deleted words from the text until only an n$^{th}$-part remain. This means that the homogeneous assumption immediately leads to a \textit{prediction} for the word-frequency of the n$^{th}$-part of the text. This prediction for Moby Dick is given in Fig. \ref{fig:3}(a). The point is that the homogeneous assumption to good approximation \textit{predicts} how the word-frequency changes when you take a part of the text. It also predicts that the power-law index $\gamma$ increases with decreasing text size. The conclusion inferred from Fig. \ref{fig:3}(a), is that the randomness implied by the homogeneous assumption gives both a qualitative and a quantitative agreement with the data (Fig. \ref{fig:A3} in Appendix \ref{apdx} gives the relative error for the predictions and compared to the same error obtained from the scaling assumption).

The second randomness assumption enters through the maximum entropy principle. This is the same principle which in physics gives rise to the ideal gas law by assuming that the collisions between the gas-molecules in a container are random or the Gauss-distribution by assuming that the deviations around some average are random. In the present context it can be formulated as the Random Group Formation (RGF) \cite{baek11}. RGF predicts the word-frequency distribution from the sole knowledge of the total number of words $M$, the number of distinct words $N$ and the frequency of the most common word $k_{max}$ \cite{baek11}. The point is that the RGF-prediction is a general prediction where the randomness is incorporated into the maximum entropy principle: it predicts the probability $P(k)$ that an object belongs to a group containing $k$ objects provided that you know that the total number of objects is $M$, the total number of groups is $N$ and that the number of objects in the largest group is $k_{max}$ \cite{baek11}. Thus the RGF-prediction involves no linguistic information other than the identification between objects and words and between groups and distinct words. What is reflected in the word-frequency distribution is some general property, which texts share with many other completely unrelated phenomena \cite{baek11,yan14,lee12,baek11b,bokma13}. Thus RGF predicts the probability $P_M(k)=N(k)/N$ for the full text without any explicit linguistic information and the homogeneity assumption transforms this expectation into the text-parts of length $M/n$ using the Random Book Transformation (RBT) inherit in the text homogeneity assumption \cite{bern09}
\begin{equation}
\label{eq_RBT1}
\textbf{P}_{M/n}(k)=B\sum_{k'=k}^M \textbf{A}_{kk'}\textbf{P}_M(k'),
\end{equation}
where $\textbf{P}_{M/n}$ and $\textbf{P}_{M}$ are column matrices corresponding to $P_{M/n}$ and $P_{M}$. The transformation matrix $\textbf{A}_{k'k}$ is given by 
\begin{equation}
\label{eq_RBT2}
\textbf{A}_{kk'}=(n-1)^{k'-k}n^{-k'}C^{k'}_k,
\end{equation}
where $C^{k'}_k$ is binomial coefficient. $B$ is given by the normalization condition
\begin{equation}
\label{eq_RBT3}
B^{-1}=\sum_k^M\sum_{k'=k}^M \textbf{A}_{k'k}\textbf{P}_M(k').
\end{equation}
Thus, by combining the maximum entropy randomness with the homogeneity randomness, the word-frequency for parts of Moby Dick can be entirely predicted from the sole knowledge of $M$, $N$ and $k_{max}$ for the full text. These predictions are given by the dashed curves in Fig. \ref{fig:3}(b). The agreement with the data (full drawn curves in Fig. \ref{fig:3}(c)) is striking (see Fig. \ref{fig:A3}(b) in Appendix \ref{apdx} for the relative errors). The fact that you to good accuracy can \textit{predict} the features of the word-frequency from two very general assumptions of randomness and without any specific linguistic information conversely suggests that you can extract basically no linguistic information from the shape of the word-frequency distribution.

\section{Shape Invariance under the Random Book Transformation} \label{sec:4}
\begin{figure}
\includegraphics[width=0.5\textwidth]{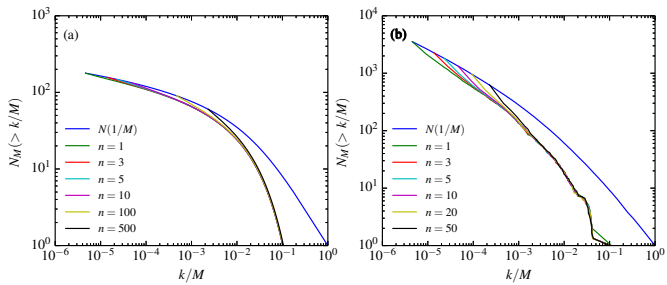}
\caption{Results for partitioning of frequency probability distributions of the form $P_M(k) \propto exp(-kb)/k^\gamma$. (a) is the exact results for the special frequency probability function $P_M(k) \propto exp(-kb)/k^1$  given by Eq. (\ref{eq:gamma1}). Here $M$ and $b$ are the same as for Moby Dick. In this case the $N_M'(\geq k/M')$-curves are almost collapsed for $M'=M/n$ with $n\leq 10$. However for $n>10$ the deviations become significant. (b) shows the case for $P_M(k) \propto exp(-kb)/k^{1.5}$ again with the same $M$ and $b$ as for Moby Dick. In this case a book consistent with the distribution is created and partitioned. One notes that the deviations are significant already for $n\geq 3$.}
\label{fig:4}
\end{figure}

The RBT-transformation given in Eqs. (\ref{eq_RBT1},\ref{eq_RBT2},\ref{eq_RBT3}) predicts how the probability distribution $P_M(k)=N(k)/N$ for the full text changes into  $P_{\frac{M}{n}}(k)$ for the n$^{th}$ part when assuming text-homogeneity. At the same time the RGF-function $P_M(k) \propto exp(-kb)/k^\gamma$ with $\gamma$ typically in the range $[1.5,2]$ gives both a qualitatively and quantitative account of word-frequency distributions in real texts \cite{baek11}. This means that one can use the functional form $P_M(k) \propto exp(-kb)/k^\gamma$ in order to understand how the general shape of the frequency distribution influences what happens when one takes an n$^{th}$-part of the text. This leads to two exact mathematical results which helps clarifying the situation. The first result is that $P_M (k) \propto exp (-kb)/k$ under the RBT transforms as
\begin{equation}
\label{eq_gamma=1}
P_{\frac{M}{n}}(k)\propto \frac{exp(-k\ln(n(e^b-1)+1))}{k} 
\label{eq:gamma1} 
\end{equation}
This means that the limit case $b=0$ and $P_M(k)\propto 1/k$ is invariant under the transformation. However, $1/k$ is not normalizable, so $b$ has to be larger than zero. If $b$ is small enough then Eq. (\ref{eq:gamma1}) reduces to  $P_{M'=M/n}(k)=A(n)exp(-knb)/k$. The average $<k>=\int_1^{\infty} P_{M'=M/n}(k)dk$ then becomes $<k>=A(n)e^{-bn}/bn$, so that $M'/N'=<k>=A(n)/bn$ for small $bn$. It follows that $M'N'P_{M'}(k)=M^2bexp(-bnk)/(kn)$. Consequently, in this special case and provided $n$ is small enough, $M'N'P_{M'}(k/M')$ obeys the scaling proposed by Font-Clos et al in Ref. \cite{font-clos2013}. Fig. \ref{fig:4}(a) shows the result for Eq. (\ref{eq:gamma1}) for the same $M$ and $b$ as for Moby Dick. As seen from the figure, in this special case of $\gamma=1$, the approximate scaling is according to the ``randomness-view" predicted to hold to good approximation down to a tenth of the original text-length. However, for larger $n$ the situation shown in Figs. \ref{fig:4}(b) and (c) is recovered, as it must from general considerations.

The second analytical solution to  Eqs. (\ref{eq_RBT1},\ref{eq_RBT2},\ref{eq_RBT3}) is $P_M(k) \propto exp(-kb)/k(k-1)$  which transforms into 
\begin{equation}
\label{eq_gamma=2}
P_{\frac{M}{n}}(k)\propto \frac{exp(-(k-1)\ln(ne^b-n+1))}{k(k-1)}  
\end{equation}
One notes that this form diverges at $k=1$. The point is that if you start with $P_M(k) \propto exp(-kb)/k^\gamma$ and $\gamma> 1$ then you approach this divergent form with increasing $n$. This tendency is illustrated in Fig. \ref{fig:4}(b) for the case $P_M(k) \propto exp(-kb)/k^{1.5}$ with the same $M$ and $b$ as for Moby Dick. In this case the deviation from the approximate scaling form is significant already for $n=3$. 
The point is that since a real text corresponds to $1.5<\gamma<2$, it follows that it will always change shape whenever $n$ becomes large enough. It also follows that the discrepancy between a scaling curve and the data for a given $n$ will depend on the starting shape. The larger $\gamma$ the starting shape has, the larger discrepancy for a given $n$. This explains why the discrepancy in case of Potter 1-7 for $n=20$ is smaller than for Moby Dick (compare Figs. \ref{fig:2}(a) and (b)). Another aspect, which is to some extent reflected in the difference of the 20$^{th}$-parts for Moby Dick and Potter 1-7, is related to the Meta-book concept discussed in Ref. \cite{bern09}. The meta-book of an author is all novels written by an author added together to a single text. The larger part of this text you analyze, the smaller the power law index $\gamma$ \cite{bern09}. It was suggested that, in the limit of an infinite text, $\gamma$ approaches 1 and the distribution $P_M(k)$ approaches the limit form $1/k$ \cite{bern09}. This means that the longer the text, the smaller the $\gamma$ and consequently also the smaller the difference in functional form, when taking an n$^{th}$-part. Thus the fact that Potter 1-7 is about five times longer than Moby Dick suggests that the discrepancy between the 20$^{th}$-part and the full text should be larger for Moby Dick than for Potter 1-7, in accordance with Figs. \ref{fig:2}(a) and (b).

In conclusion we find that an approximate scaling form in a special case indeed emerges from the randomness assumption that words with a given frequency are equally likely to appear anywhere in a text.

\section{Concluding Remarks} \label{sec:5}
This paper discusses both general and specific issues in connection with analyzing word-frequency distributions. The first general issue concerns the two seemingly not compatible views on word-frequencies i.e. the `Zipf's view' attributing specific properties to the distribution and the `Randomness view' which emphasizes the non-specific character of the distribution. We have here shown that the `Randomness view' \textit{predicts} word-frequency distributions both qualitatively and quantitatively. The conclusion drawn from this is that the shapes of word-frequency-distributions are general features, which word-distribution share with a multitude of totally unrelated phenomena and that in fact the linguistic content, which can be drawn from these distribution, is basically nil. This is contrary to the `Zipf's view', which often presumes that the shapes of  word-frequency distributions carry specific linguistic information. As a concrete example we discussed a generalized scaling which could well have reflected such a specific feature \cite{font-clos2013}. However carefully analysis showed that this particular scaling is not borne out by the data and does not suggest any specific feature beyond the randomness.  The second general issue is the conceptual difference between \textit{predicting} and \textit{fitting}: The `Randomness view' \textit{predicts} word-frequency distributions from general assumptions, whereas \textit{fitting} to particular curve-forms is a common procedure within this field and a successful fit to the data over a limited range is often furthermore taken as evidence of the correctness of the assumptions underlying the curve-form. In the concrete example discussed here, it was proposed that the suggested scaling function could be \textit{approximately }parameterized by a single curve-form with two free parameters \cite{font-clos2013}. However, as shown, such a fitting does not imply anything beyond what is contained in the randomness-view.

More generally from the point of the present work, an analysis based on fitting word-frequency data to power-laws has but little information value, in spite of its long history stemming from Zipf's early work in 1932 \cite{zipf32}. The questions of how a power-law fitted to data for $N(M)$ (usually called Heaps' law) relates to a power-law fitted to word-frequency distribution $P(k)$ (usually called Zipf's law) is discussed in Ref. \cite{lu10}. From the randomness-perspective such power-law fittings carries no additional information: $N(M)$ and $P(k)$ carry the same random-information and the $N(M)$ can be directly obtained from $P(k)$ \cite{bern11b,lu10}. The relation between Heaps' law and Zipf's law in the present context is further illustrated in Appendix \ref{apdx}. Another general point is that that $P(k)$ is never quite a power-law as can be seen from Fig. \ref{fig:1}(b) and (c) and Fig. \ref{fig:A1} in Appendix \ref{apdx}. The $P(k)$ in a log-log plot bends for higher values of $k$ in accordance with RGF-form from the maximum entropy principle. Sometimes this bend is, in accordance with the scale-freeness ideas, associated with yet a second power-law as in the case of language corpora in Ref. \cite{petersen12}. In Appendix \ref{apdx} we show that such corpora are also well described by the randomness-view and the RGF-form (compare Fig. \ref{fig:A2}). Thus fitting the data by two power-laws does by itself not imply any specific language feature. 

Word-frequencies are not the only example for which outcomes for some complex deterministic systems reduce to random distributions. Other examples are e.g. distribution of species into taxa in biology \cite{bokma13}, chemical reaction networks \cite{lee12}, family names \cite{baek11,baek11b}, sizes of population centers \cite{baek11} and travel distance distributions \cite{baz10,yan13}. Thus the question of random versus specific is relevant in a much broader context than just word-frequencies.

\textbf{Acknowledgments:}
Supported by NSFC under the grant No. 61304177 and the Fundamental Research Funds of BJTU under the grant No. 2015RC042.


\appendix
\renewcommand\thefigure{A\arabic{figure}}
\setcounter{figure}{0}

\begin{figure*}
\includegraphics[width=0.8\textwidth]{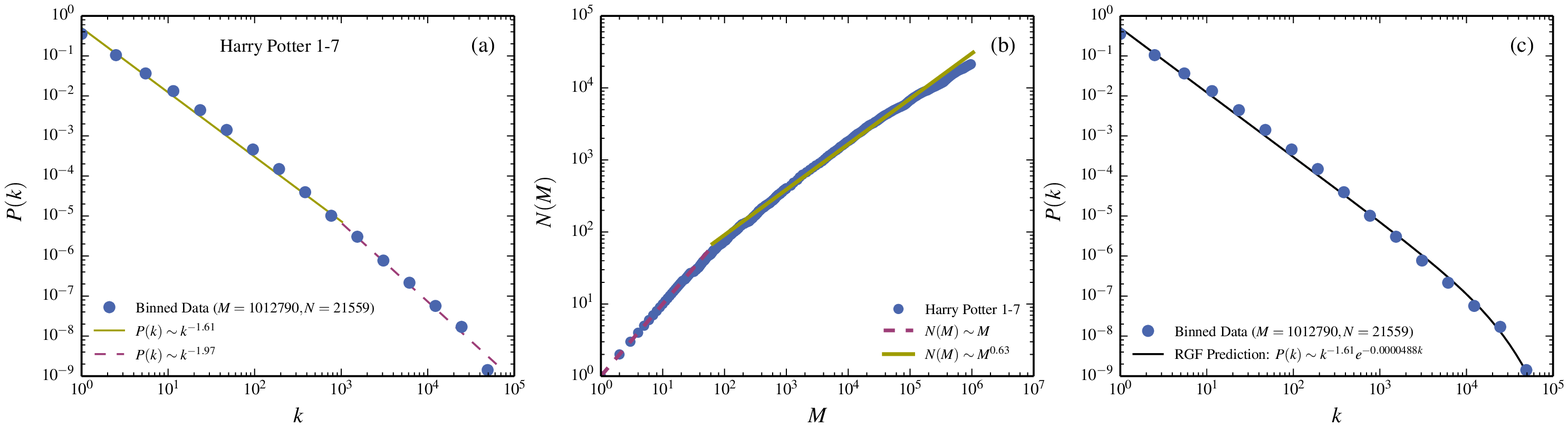}
\caption{Power-law fitting versus randomness predictions. The data for the Potter text (same as in Fig. \ref{fig:1} (c)) are replotted as $P(k)$ versus $k$ in (a) and in (b) as $N(M)$ versus $M$. In these log-log plots a power-law corresponds to a straight line. A straight line approximation in (a) is called a Zipf's law approximation and in (b) a Heaps law approximation. Since the data is curved you can get a rather good representation by \textit{fitting} two straight lines. The left of these is in (a) usually referred to as the Zipf's law approximation and the right either as a cut-off or a second  scale-free region. In (b) the right straight-line is referred as the Heaps law approximation and the left as the asymptotic limit where each word is a new word. (c) shows that you can \textit{predict} the data to good approximation directly from the sole knowledge of $M$, $N$ and $k_{max}$.}

\label{fig:A1}
\end{figure*}

\section{}\label{apdx}
\begin{figure}
\includegraphics[width=0.5\textwidth]{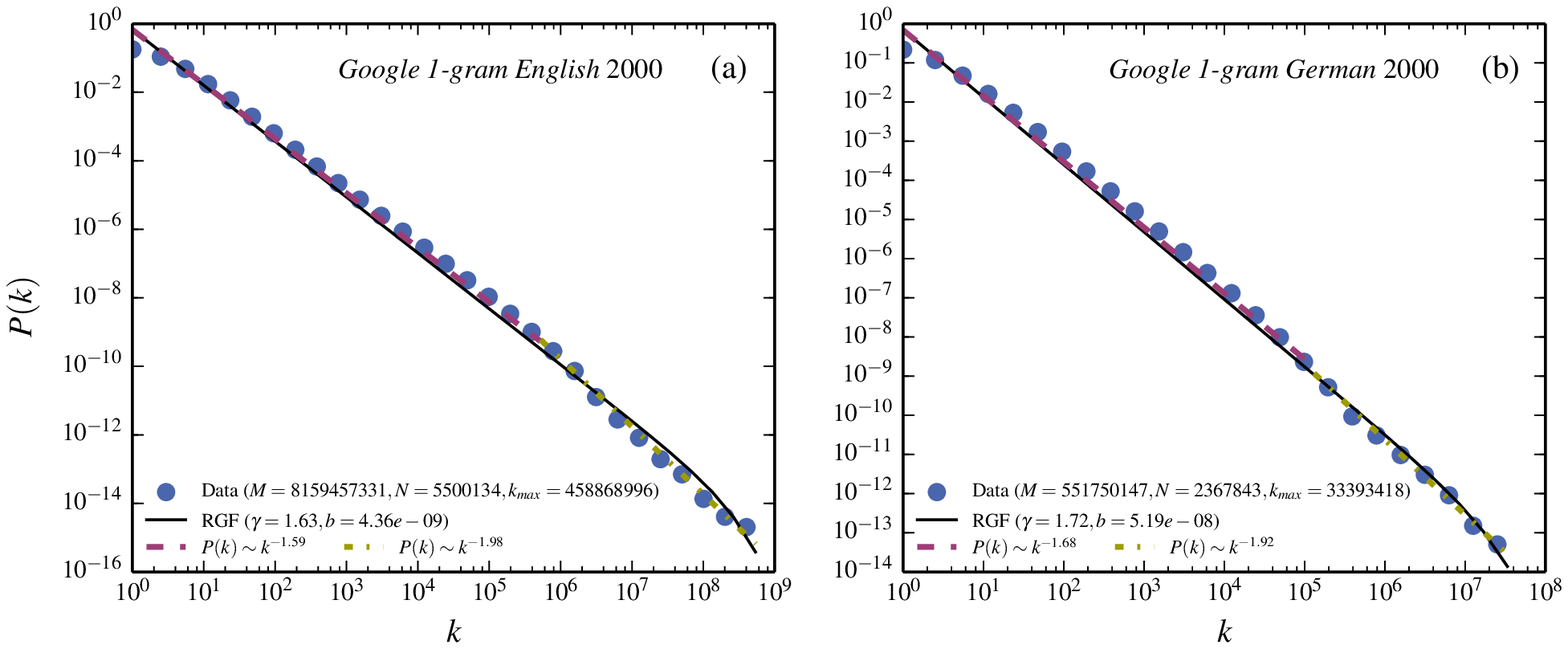}
\caption{Power-law fitting versus randomness predictions for Google 1-grams. (a) is for English and (b) for German. In both cases you can {\textit{fit}} the data for $Pk)$ versus $k$ in a log-log plot with two straight lines. However, the data is already well \textit{predicted} from the sole knowledge of $M(N)$, $N$ and $ k_{max}$ using the randomness property.}    

\label{fig:A2}
\end{figure}

Zipf's law is often expressed in terms of the probability function $P(k)=N_M(k)/N$ where $N=N_M(\geq 1)$ is the total number of specific words. A log-log plot of $P(k)$ is given  Fig. \ref{fig:A1}(a). The data is the same Potter text as in Fig. \ref{fig:1}(a) and is represented by dots in binned form in the figure. Zipf's law means, in its original form, that this data should fall on a straight line with slope -2. As seen in the figure the left part of the data can we well approximated with straight line of slope -1.6 (full line in Fig. \ref{fig:A1}(a)). This is often taken as evidence that a language can be associated with the specific property of scale-freeness, although with a different exponent than the -2 stipulated by Zipf's law. However, since the data actually follows a bent curve the right part of the data fits better to a straight line with a steeper slope (dashed line in the Fig. \ref{fig:A1}(a)). This means that the data can be rather well \textit{fitted} by two straight-lines of different slopes. The fact that you need two straight-lines to fit the data was discussed in Ref. \cite{petersen12} in case of large language corpora. Two examples from the Google 1-grams data \cite{google} are shown in Fig. \ref{fig:A2}. As seen both sets can be well represented by \textit{fitting} to two straight lines in accordance with Ref. \cite{petersen12}. The full drawn curve in Figs. \ref{fig:A2} (a) and (b) are \textit{predictions} from the randomness view. The randomness prediction maintains that if you know $M$, $N$,and $k_{max}$ you can regardless of language predict $P(k)$ very well. Furthermore the randomness predictions can also be well \textit{fitted }to two straight-lines. Thus there is little evidence that you can extract any additional information by $fitting$ to two straight-lines than what is already contained in the randomness view. Fig. \ref{fig:A1} (c) gives the corresponding randomness prediction of the Potter text.

\begin{figure}
\includegraphics[width=0.5\textwidth]{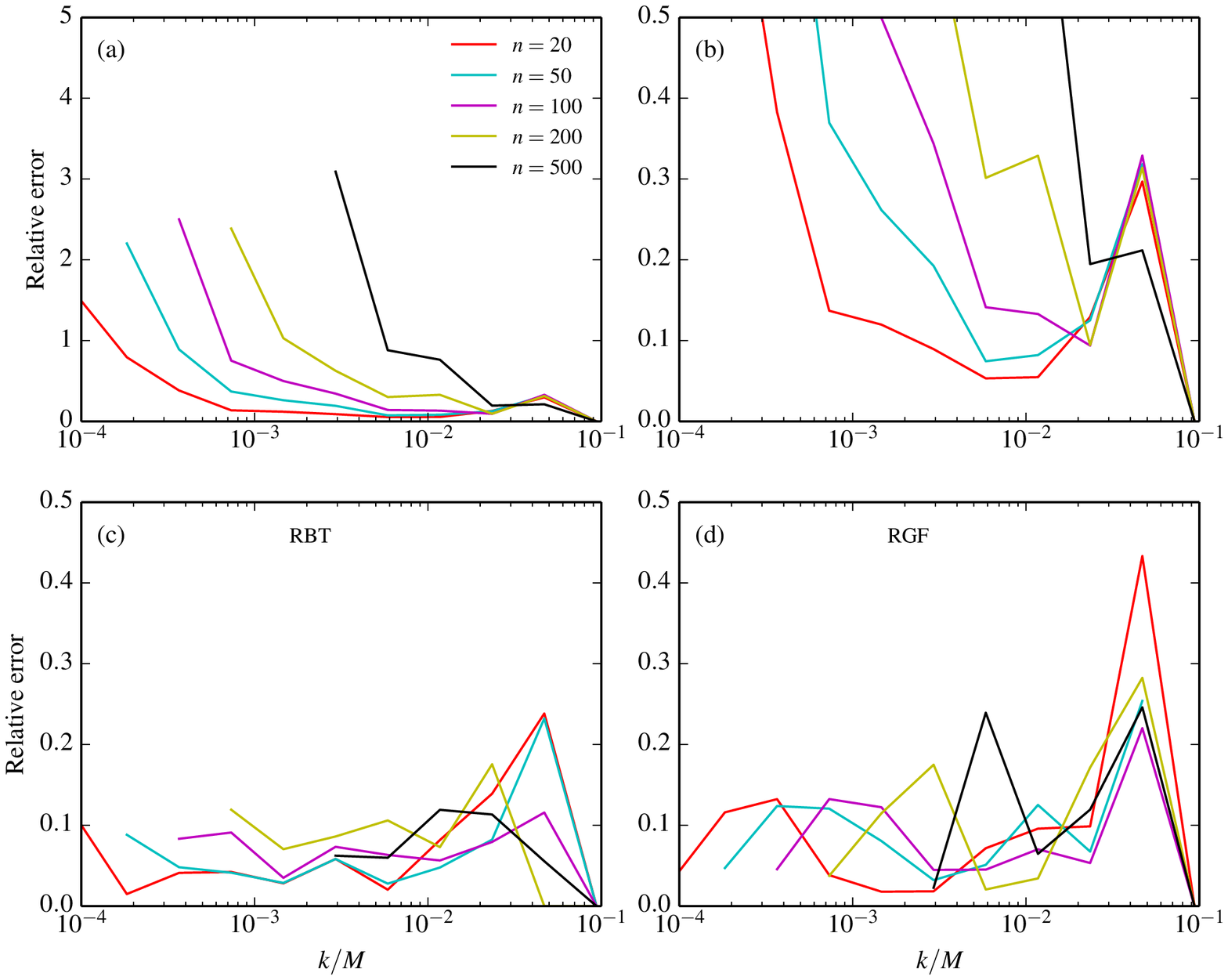}
\caption{Relative error for the predictions illustrated in Figs. \ref{fig:2}(a), \ref{fig:3}(a) and  \ref{fig:3}(b). The relative error is given by
 {$\sqrt{(k/k_{pred}-1)^2}$} where $k$($k_{pred}$) is the measured (predicted) value, respectively. (a) is the scaling prediction that the data for the n$^{th}$-part should fall on on the full text data n$^{th}=1$. The error becomes large for towards smaller $k$ values (upto 200-300\%) and progressively worse with decreasing text-length. (b) A blow up showing where the error becomes larger than 50\% for the various text-lengths. (c) shows the randomness prediction which has much smaller relative errors in the range of 10\% over the complete range. (d) is the prediction based on the sole knowledge of $M$, $N$ , and $k_{max}$ for the complete text. The relative errors are only slightly larger than for (c). It is the small relative errors in (c) and (d) which are reflected in the near overlaps between prediction and data in Fig. \ref{fig:3}.}    
\label{fig:A3}
\end{figure}

Figure \ref{fig:A1} (b) shows $N(M)$ for the Potter text i.e. the number of specific words as a function of the length of the text. Also this curve can to some extent be \textit{fitted} to a power law (straight full line in Fig. \ref{fig:A1} (b)). This is in approximate agreement with Heaps law which states that $N(M)$ has a power-law character. However, since the first words are specific the slope of the power-law is $\approx 1$  in the left-most part and it then decreases because the curve is apparently bent. From the random-view perspective, $P(k)$ and $N(M)$ are simply related and contain the same information \cite{bern11b}. In particular the randomness view predicts that if $P(k)$ is a power-law then the corresponding Heaps law can only be approximate \cite{bern11b}. Nevertheless  one can derive approximate relations between the approximate power-laws \cite{lu10}. In the simplest approximation it is just $\alpha=\gamma -1$ provided $P(k)\propto 1/k^\gamma$ and $N(M)\propto M^\alpha$. This relation holds pretty well for the full drawn lines in Fig. \ref{fig:A1} (a) and (b). In Fig. \ref{fig:2} (c) and (d) we have in, correspondence with Fig. \ref{fig:A1} (a), used the left part of the data to extract an approximate power-law.

Figure \ref{fig:A3} supplies a measure of the precision in Figs. \ref{fig:2}(a), \ref{fig:3}(a), and \ref{fig:3}(b). Fig. \ref{fig:2} (a) compares $n$-parts of Moby Dick with the full text $n^{th}=1$. The relative error made is given in Figs. \ref{fig:A3}(a) and (b): Fig. \ref{fig:A3}(a) shows that the relative error becomes very large (upto 300\%) for the smaller values. Fig. \ref{fig:A3}(b) shows the error for larger values in a blown up scale. This should then be compared to the relative error between the data and the randomness prediction shown in Fig. \ref{fig:A3}(c) which corresponds to Fig. \ref{fig:3}(a). By comparing Figs. \ref{fig:A3}(a), (b) and (c), one finds that the randomness prediction is a better prediction than the scaling prediction. Fig. \ref{fig:3}(c) corresponds to prediction you get from the sole knowledge of $M$, $N$ and $k_{max}$ for the full Moby Dick. Thus without any knowledge of the author or what language the text is written in. From this point of view the smallness in the relative error is notable.

\end{document}